\documentclass[aps,prb,twocolumn,superscriptaddress]{revtex4-2} 
\usepackage{graphicx}
\usepackage{xcolor}
\usepackage{subfigure}
\usepackage{amsmath}
\usepackage{epsfig}
\usepackage{chemfig}
\usepackage{braket}
\usepackage{chemfig}
\usepackage{mathtools}
\usepackage{bm} 
\usepackage[colorlinks= true,urlcolor=blue,linkcolor= blue,citecolor=blue,bookmarks=false,pdfstartview=]{hyperref}

\begin{document}

\setchemfig{atom sep=2em}

\title{Heterometallic spin-1/2 quantum magnet under hydrostatic pressure}%

\author{M.~J.~Coak}
\affiliation{Department of Physics, University of Warwick, Gibbet Hill Road, Coventry, CV4 7AL, UK}
\affiliation{School of Physics \& Astronomy, University of Birmingham, Edgbaston, Birmingham, B15 2TT, UK}
\author {D. Kamenskyi}
\affiliation{Department of Physics, Humboldt-Universit\"at zu Berlin, Berlin, Germany}
\affiliation {Experimental Physics V, Center for Electronic Correlations and Magnetism, Institute of Physics, University of Augsburg, 86159 Augsburg, Germany}
\affiliation {Molecular Photoscience Research Center, Kobe University, Hyogo 657-8501 Kobe, Japan}
\affiliation {Institute of Space Research, German Aerospace Center (DLR), 12489 Berlin, Germany}
\author{S.~P.~M.~Curley}
\affiliation{Department of Physics, University of Warwick, Gibbet Hill Road, Coventry, CV4 7AL, UK}
\author{B.~M.~Huddart}\thanks{Current address: Clarendon Laboratory, Department of Physics, University of Oxford, Parks Road, Oxford OX1 3PU, United Kingdom}
\affiliation{Department of Physics, Centre for Materials Physics, Durham University, Durham, DH1 3LE, United Kingdom}
\author{J.~P.~Tidey}
\affiliation{Department of Chemistry, University of Warwick, Gibbet Hill, Coventry CV4 7AL, U.K}
\author {A. Chmeruk}
\affiliation{Theoretische Physik III, Center for Electronic Correlations and Magnetism, Institute of Physics, University of Augsburg, 86135 Augsburg, Germany}
\affiliation{Augsburg Center for Innovative Technologies (ACIT), University of Augsburg, 86135 Augsburg, Germany}
\author {T. Sakurai}
\affiliation {Core Facility Center, The University of Osaka, Yamadaoka 2-1, Suita, Osaka 565-0871, Japan}
\author {S. Okubo}
\author {H. Ohta}
\affiliation {Molecular Photoscience Research Center, Kobe University, Hyogo 657-8501 Kobe, Japan}
\author {S. Kimura}
\author {H. Nojiri}
\affiliation {Institute for Materials Research, Tohoku University, Katahira 2-1-1, Sendai, Miyagi 980-8577, Japan}
\author{D.~Graf}
\affiliation{National High Magnetic Field Laboratory, Florida State University, Tallahassee, Florida 32310, USA}
\author{S.~J.~Clark}
\affiliation{Department of Physics, Centre for Materials Physics, Durham University, Durham, DH1 3LE, United Kingdom}
\author{Z.~E.~Manson}
\affiliation{Department of Chemistry and Biochemistry, Eastern Washington University, Cheney, Washington 99004, USA}
\author{J.~L.~Manson}
\thanks{Deceased.}
\affiliation{Department of Chemistry and Biochemistry, Eastern Washington University, Cheney, Washington 99004, USA}
\author{T.~Lancaster}
\affiliation{Department of Physics, Centre for Materials Physics, Durham University, Durham, DH1 3LE, United Kingdom}
\author{P.~A.~Goddard}
 \email{p.goddard@warwick.ac.uk}
\affiliation{Department of Physics, University of Warwick, Gibbet Hill Road, Coventry, CV4 7AL, UK}

\begin{abstract}

We investigate the properties of CuVOF$_4$(H$_2$O)$_6$$\cdot$H$_2$O, in which two different spin species, Cu(II) and V(IV), form antiferromagnetic spin-1/2 dimers with weak interdimer coupling provided via hydrogen bonding. Using radio-frequency susceptometry and electron-spin resonance (ESR), we show how the temperature-magnetic field spin-dimer phase diagram evolves as a function of applied hydrostatic pressure and correlate this with pressure-induced changes to the crystal structure. These results, coupled with pressure-tuned DFT calculations, confirm the prior prediction that the primary exchange interaction is mediated via an unusual mechanism in which the V(IV) ions provide considerable spin density to the oxygen that joins the two spins in each dimer and which lies along the Jahn-Teller axis of the Cu(II) ion. In addition, the dissimilarity in the spins that make up each dimer unit leads to a non-linear field dependence of the electronic energy levels as detected in the ESR measurements.  

\end{abstract}

\maketitle

\section{Introduction}

Symmetry breaking is an important concept throughout the natural world encompassing, among other things, issues related to the origins of matter, mass and life~\cite{Krakauer2023,Canetti2012,Higgs1964,Sandars2005}. In condensed matter physics the concept can be split into two categories: spontaneously broken symmetry, which occurs when the ground state of the system breaks a symmetry that is respected by the Hamiltonian, such as translational symmetry in crystals, rotational symmetry in ferromagnets and global phase symmetry in superconductors~\cite{Anderson1964, Landau1937, Weinberg1986}; and explicitly broken symmetry in which small perturbative terms in the Hamiltonian themselves break the symmetry in question.  

At low temperatures, arrays of weakly coupled Heisenberg spin-1/2 antiferromagnetic dimers adopt a quantum disordered state of spin singlets, which is separated by an energy gap from the higher lying triplet levels. A magnetic-field-induced phase transition takes place when the spins spontaneously break rotational symmetry and form a state of long-range $XY$ magnetic order. This transition can be mapped on to the condensation of a system of bosonic or triplon excitations, whose masses can be strongly renormalized by the effect of quantum fluctuations~\cite{Zapf2014a,Oosawa1999,Ruegg2003,Jaime2004,Coak2023}.   

CuVOF$_4$(H$_2$O)$_6$$\cdot$H$_2$O is a spin-1/2 dimer material, with weak interdimer coupling provided via hydrogen bonding. At ambient pressure the system displays the typical temperature--field phase diagram, hosting quantum-disordered, long-range $XY$-ordered and field-aligned magnetic states~\cite{Curley2021a}.  However, because the spin dimers are comprised of two dissimilar $S = 1/2$ ions, Cu(II) and V(IV), the magnetic Hamiltonian explicitly breaks rotational symmetry in a very distinct way. Two consequences of this have previously been observed.  First, the local environments around the dimerized ions do not share the same local coordinate system. This gives rise to the possibility of an anisotropic Dzyaloshinskii- Moriya (DM) interaction which can lead to the appearance of otherwise forbidden transitions ($\Delta S = \pm 1$) in electron-spin resonance (ESR) measurements ~\cite{Matsumoto2008}. Another potential cause of these transitions is the difference in g-factors between the Cu and V ions that comprise the dimer unit~\cite{Abragam1970}. Here, using the ESR measurements, we determine the individual g-factors of the Cu(II) and V(IV) ions. Second, the primary intradimer exchange interaction is predicted to be mediated via an unusual mechanism in which the V(IV) ions provide considerable spin density to the oxygen that joins the two spins in each dimer and which lies along the Jahn-Teller axis of the Cu(II) ion~\cite{Curley2021a}. 

Here, by performing ESR, bulk susceptometry, X-ray diffraction and density functional theory (DFT) calculations under applied pressure we show that the primary intradimer interaction strengthens with pressure up to a structural phase transition without significant change in interdimer coupling. At the structural transition, the local spin environments rotate dramatically with respect to one another, disrupting the H-bonded network that holds the dimers together, but affecting the Cu–O–V bond only marginally. The relatively small change in magnetic properties elucidated from ESR across the transition confirms that it is indeed the Cu–O–V bond that supports the primary interaction pathway and this conclusion is supported by the DFT results. The ESR spectra reveal an enhancement of the intradimer exchange interaction and modification of the $g$-factors of Cu(II) and V(IV) ions upon applying hydrostatic pressure. Deviations from the linearity in the frequency-field dependence of the magnetic excitations allow us to disentangle the $g$-factors of the Cu(II) and V(IV) ions.

\section{Methods}

All chemical reagents for the sample synthesis were obtained from commercial sources and used as received. CuF$_2$ ($0.4579$\,g, $4.51$\,mmol) and VF$_4$ ($0.5724$\,g, $4.51$\,mmol) were dissolved in HF in separate plastic beakers and then combined. 2,5-methylpyrazine ($0.975$\,g, $9.02$\,mmol) was added to the reaction, which was filtered twice to yield a clear green solution. Slow evaporation at room temperature yielded rectangular blue crystals around 1--4\,mm on an edge. Note: hydrofluoric acid is extremely dangerous and must be handled with care, sufficient training and appropriate personal protective equipment.   

Pressure dependent single-crystal X-ray diffraction measurements were performed at the University of Warwick using a Diacell TozerDAC turn-buckle diamond anvil cell (DAC) mounted onto a Rigaku Synergy XtaLAB Synergy-S diffractometer, equipped with a HyPix photon-counting hybrid pixel array area detector. Measurements were performed using Mo K$_{\alpha}$  radiation (0.71073 \r{A}) from a PhotonJet micro-focus sealed tube source, allowing data to be collected to standard resolution. In situ pressure was determined by tracking the pressure-dependence of the fluorescence spectrum of a small ruby chip within the DAC sample chamber. Glycerol was used as the pressure transmitting medium, which has a hydrostatic limit of 5.5 GPa \cite{Klotz2012} and can hence be considered hydrostatic across the entire range of the study.

Pressure-dependent radio-frequency (RF) susceptometry measurements were performed on single-crystals of CuVOF$_4$(H$_2$O)$_6\cdot$H$_2$O using a piston-cylinder cell with fields attained using a 36\,T resistive magnet and 18\,T superconducting magnet at the National High Magnetic Field Laboratory, Tallahasee. The pressure transmitting medium was Daphne oil 7373, which, as with the glycerol used in the X-ray measurements, provides good hydrostatic conditions in this pressure range \cite{Klotz2009}. A ruby chip located on the end of a fiber optic cable situated next to the sample provided in-situ pressure calibration. 

ESR experiments were performed using three transmission-type, multi-frequency ESR spectrometers. Measurements at ambient pressure and magnetic fields up to 15\,T have been performed using pulsed magnetic fields with a pulse duration of 10 msec \cite{Ohta2006}. Experiments at hydrostatic pressure up to 2.45(5)\,GPa and radiation frequencies up to 500\,GHz were made using a 10\,T superconducting magnet at Kobe University~\cite{Sakurai2007,Okuto2019}. In both of these experiments backward wave oscillators and Gunn diodes were employed as radiation sources. A similar pressure cell was used for high-field ESR experiments up to 25\,T  with a cryogen-free superconducting magnet at the High Field Laboratory for Superconducting Materials (Tohoku University, Japan)~\cite{Sakurai2018}. In all ESR measurements, 2,2-Diphenyl-1-picrylhydrazyl (DPPH) was used as a standard magnetic field marker.

DFT calculations were performed using the approach described in Ref.~\cite{Curley2021a}.

\section{Results}

\begin{figure}[t!]
\centering
\includegraphics[width= \linewidth]{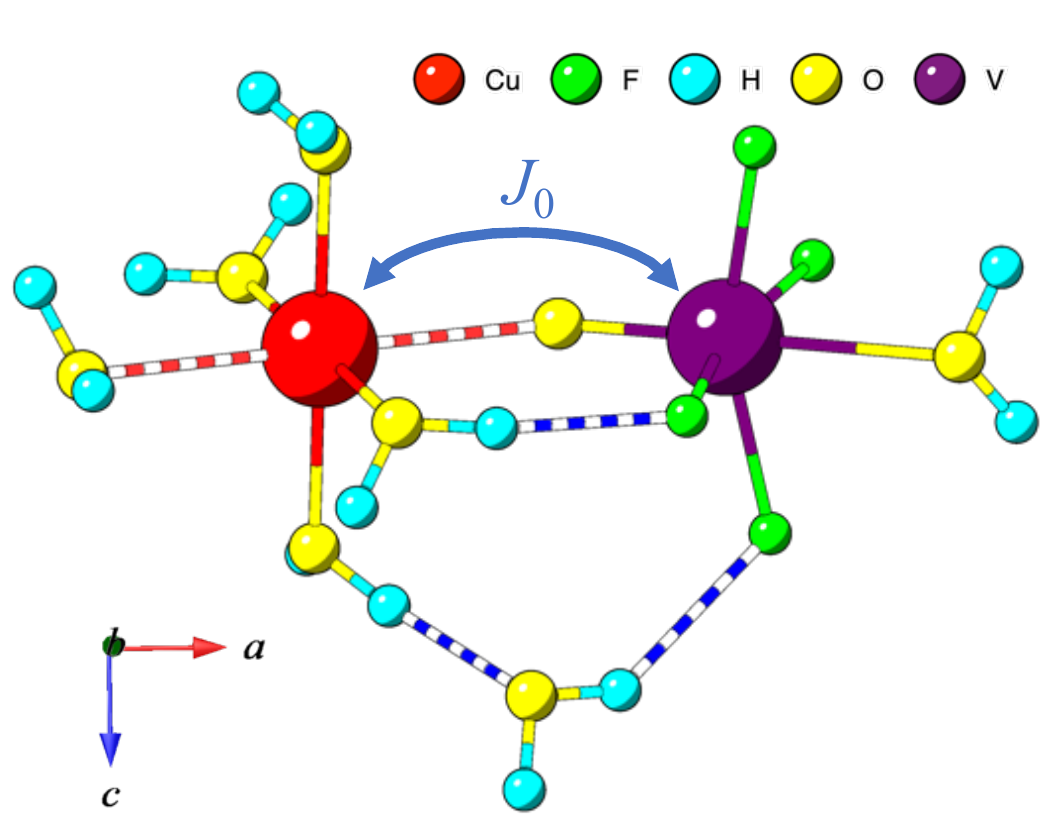}
\caption[width= \linewidth]{\small Local structure of Cu(II)–V(IV) dimer units of CuVOF$_4$(H$_2$O)$_6 \cdot$H$_2$O measured at 150\,K using X-ray diffraction. Red striped bonds indicate the Jahn-Teller axis of the Cu(II) ion, while blue stripes highlight the intradimer hydrogen bonds. The intradimer exchange $J_0$ is predicted to be mediated through the central oxygen atom via an atypical exchange mechanism~\cite{Curley2021a}.}  \label{Fig:CuV_neutron}
\vspace{-0cm}
\end{figure}

\subsection{Crystallography}
\label{sec:CuV_struc_Pdep}

The crystal structure of CuVOF$_4$(H$_2$O)$_6 \cdot$H$_2$O at ambient pressure has been discussed in detail in Refs~\cite{Donakowski2012,Curley2021a}. Figure~\ref{Fig:CuV_neutron} shows the local structure around the spin-1/2 Cu(II)–V(IV) dimer units measured at 150\,K at ambient pressure. At the core of the spin dimers is the vanadyl VO$^{2+}$ ion, whose oxygen forms a covalent bond with Cu(II) along its Jahn-Teller (JT) elongated axis.  Previous DFT calculations have indicated that the primary intradimer exchange interaction ($J_0 = 21.6(2)$\,K, determined from susceptometry) is mediated through the vanadyl oxygen~\cite{Curley2021a}. 

\begin{figure*}
\centering
\includegraphics[width= 0.95\linewidth]{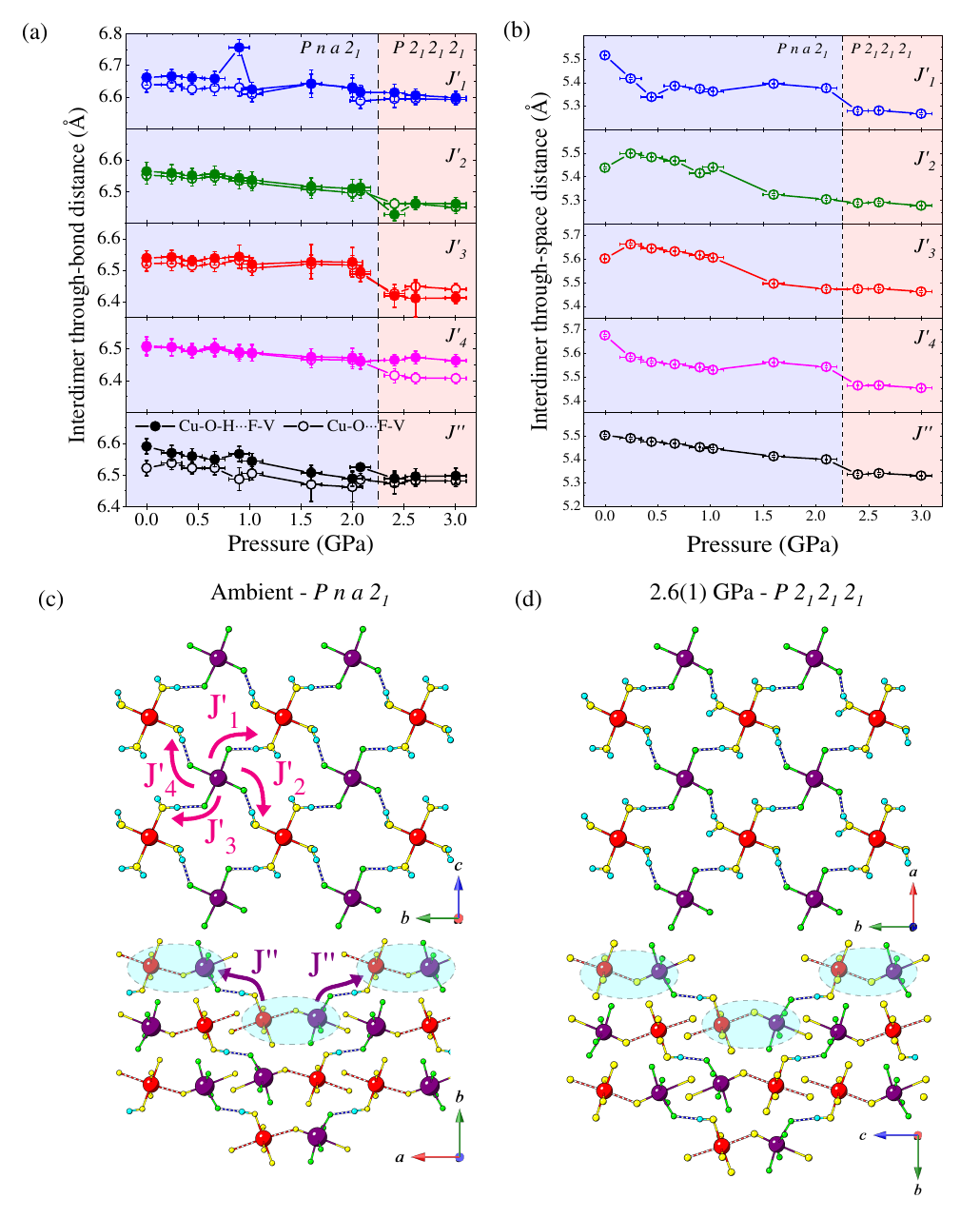}
\caption[width= \linewidth]{\small (a) Pressure dependence of the interdimer exchange pathway lengths; closed circles are Cu---O---H$\,\cdots$F---V and open circles are  Cu---O$\,\cdots$F---V bond distances, i.e. omitting the H atom (see text). (b) Pressure dependence of the through-space interdimer Cu and V distances. (c) Interdimer H-bond exchange pathway geometry in the low-pressure phase. (d) Same network for the high-pressure phase, where it can be seen to be qualitatively unchanged. Measurements are made at 300~K and errors plotted at 3$\sigma$.}  \label{Fig:CuV_inter_pdep}
\vspace{-0cm}
\end{figure*}

This is a highly unusual situation. It is known that Cu(II) JT bonds do not support significant exchange interactions, because the spin-density from the magnetic $d_{\rm x^2-y^2}$ orbitals is concentrated in the plane perpendicular to this bond. In the present case, the DFT calculations suggest that the significant $\pi$-bonding inherent in the vandyl ion allows for a sizable spin density to be located on the vanadyl oxygen which then supports a healthy superexchange~\cite{Curley2021a}. This arrangement appears to be extremely rare among Cu and V containing compounds. In a recent study, Wang {\it et al.} surveyed all 522 such materials listed on the Cambridge Structural Database and found no other examples of superexchange mediated from Cu(II) to V(IV) via a single vanadyl oxygen bridge~\cite{Wang2022}. As can be seen from Fig.~\ref{Fig:CuV_neutron}, the dimer unit also contains multiple hydrogen bonds. Such bonds are known to mediate superexchange in other materials~\cite{Goddard2008b}, and so the existence of the unusual \chemfig{Cu-O=V} exchange pathway would benefit from further experimental verification. 
As discussed in Ref.~\cite{Curley2021a}, other hydrogen bonds create Cu---O---H$\,\cdots$F---V pathways [which we label as $J'$ in Fig.~\ref{Fig:CuV_inter_pdep}(c)] linking the dimers together to form weakly coupled arrays in the $bc$ plane. Additional H-bond pathways connect neighboring dimers within the $ab$ planes. DFT results (see Ref~\cite{Curley2021a} and later) suggest that the exchange coupling via these $J''$ pathways is negligible.

The crystal structure leads to the following magnetic Hamiltonian~\cite{Curley2021a},
\begin{multline}\label{eq:Hamiltonian}
\mathcal{H} = J_0\sum\limits_{i} {\bf{\hat{S}}}_{1,i}\cdot{\bf{\hat{S}}}_{2,i}
+ \sum\limits_{<mnij>}J'_{mnij}
{\bf{\hat{S}}}_{m,i}\cdot{\bf{\hat{S}}}_{n,j} \\
- g\mu_{\rm{B}}\mu_{0}H\sum \limits_{i} {\hat{S}}^{z}_{\textit{m,i}}
\end{multline}
where $m,n$ = 1,2 label the two magnetic sites within each dimer and $i$ and $j$ denote neighboring dimers. The first two exchange terms represent, respectively, the intradimer exchange and the interdimer exchange through the hydrogen-bonded pathways. There is also the possibility of an additional DM interaction term in the Hamiltonian of the form ${\boldsymbol{D}} \cdot ({\bf{S}}_1\times {\bf{S}}_2)$, but this is expected to be small~\cite{Curley2021a}.

\begin{figure}[t!]
\centering
\includegraphics[width= \linewidth]{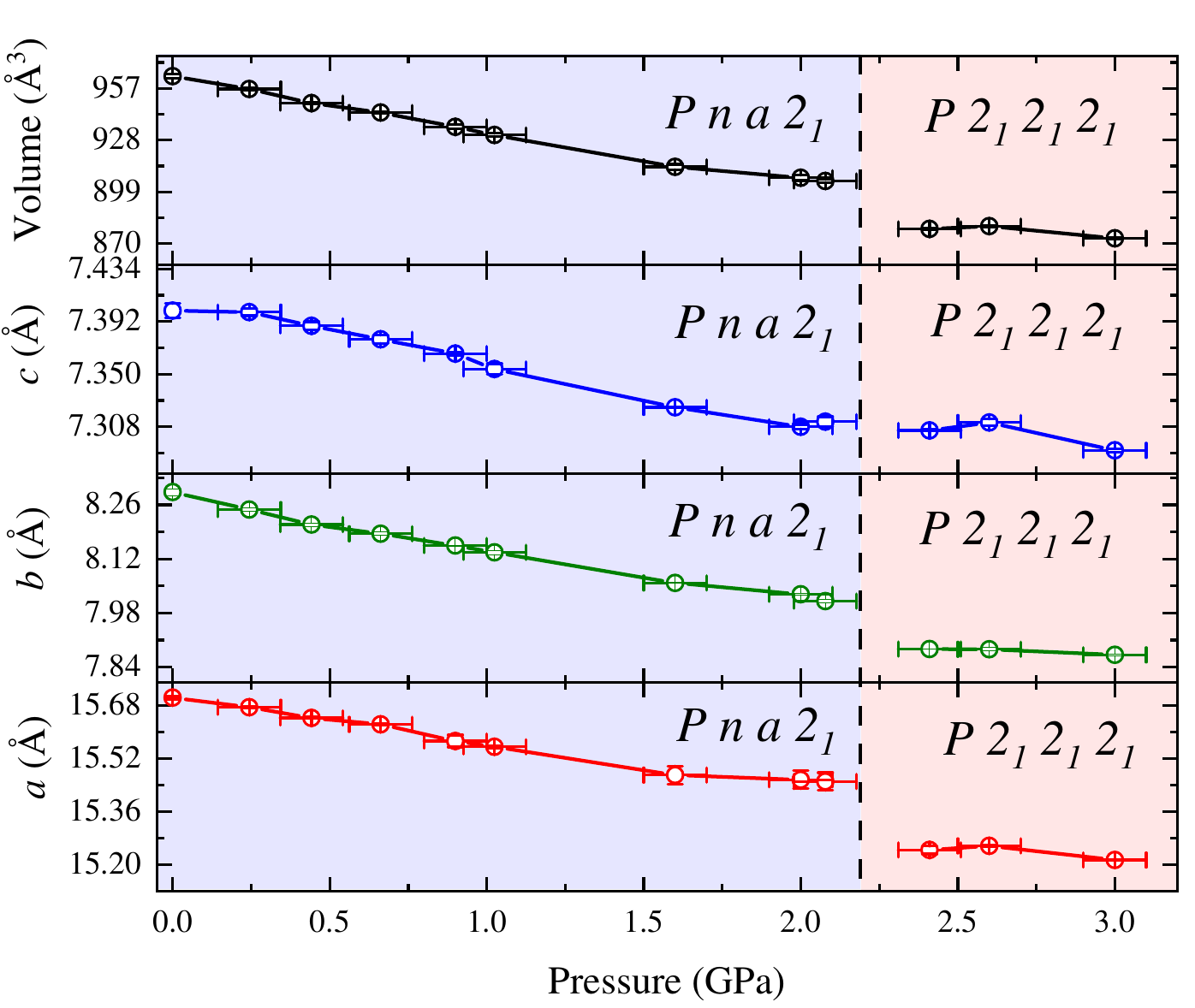}
\caption[width= \linewidth]{\small Pressure dependence of the unit cell of CuVOF$_4$(H$_2$O)$_6 \cdot$H$_2$O at 300~K with errors plotted at 3$\sigma$. Shaded regions highlight the low pressure $P n a 2_1$ and high-pressure $P 2_1 2_1 2_1$ phase, with the two regions delineated by a dashed black line.}  \label{Fig:CuV_cell_pdep}
\vspace{-0cm}
\end{figure}

Figure~\ref{Fig:CuV_cell_pdep} shows the effect of increasing hydrostatic pressure on the unit-cell of CuVOF$_4$(H$_2$O)$_6 \cdot$H$_2$O up to 3.0~GPa. Below 2.1(1)~GPa, the unit-cell volume compresses smoothly with no jumps or changes in slope that would suggest abrupt changes in structure. Within the pressure-region $2.1 < P \leq 2.4$~GPa however, the system undergoes a structural phase-transition from the polar space group $P n a 2_1$ to the chiral space group $P 2_1 2_1 2_1$, confirmed by the systematic absences of Bragg peaks in the X-ray data at pressures $P \geq 2.4(1)$~GPa.

\begin{figure}[t!]
\centering
\includegraphics[width= \linewidth]{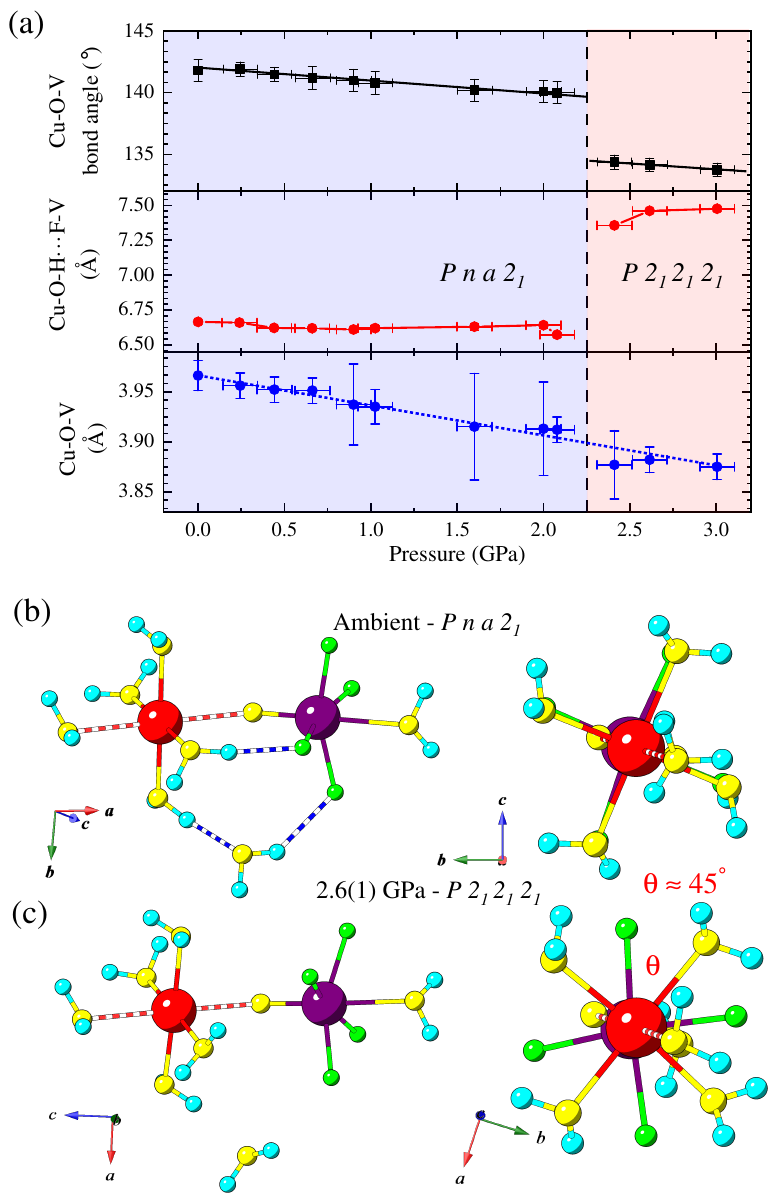}
\caption[width= \linewidth]{\small (a) Pressure dependence of the intradimer Cu---O---V summed bond distance (bottom panel), intradimer Cu---O---H$\cdots$F---V bond distance (middle panel) and Cu---O---V bond angle (top panel). Structure of the dimer-unit in the low-pressure (b) and high-pressure (c) phase. Measurements made at 300~K and errors plotted at 3$\sigma$.}  \label{Fig:CuV_intra_pdep}
\vspace{-0cm}
\end{figure}

The structural effect of pressure can be investigated further by first looking at the change in the interdimer pathways. Figure~\ref{Fig:CuV_inter_pdep}(a) shows that, within the $P n a 2_1$ phase, the interdimer Cu---O---H$\,\cdots$F---V through-bond distances (filled circles) do not decrease significantly (i.e., within 3$\sigma$ confidence, depicted by the error bars) upon increasing pressure, contrary to what one might expect.
The true position of the electron deficient hydrogens is typically poorly defined in X-ray experiments, an issue compounded when using a pressure cell, where a significant portion of reciprocal space is obscured by the cell body. These are accordingly treated as geometrically constrained, rotating groups at standard distances, yet refine to reasonably consistent positions as seen in the geometric trends, allowing their fair assessment. To assess the effect of this issue on our conclusions, Figure~\ref{Fig:CuV_inter_pdep}(a) also shows the pressure dependence of the summed direct Cu---O$\,\cdots$F---V bond distances (open circles), which are seen to agree well with the Cu---O---H$\,\cdots$F---V summed bond distances, indicating that, as expected, O---H bonds point almost directly at the highly electronegative F ligands.

By contrast to the through-bond interdimer exchange pathways, Figure~\ref{Fig:CuV_inter_pdep}(b) shows that the through-space distances between Cu and V ions on adjacent dimer-units do clearly decrease upon increasing pressure.
This comes about because, as pressure brings neighboring dimer-units into closer proximity, the increased repulsion between the dimers induces rotations of the water ligands. This leaves the interdimer Cu---O---H$\,\cdots$F---V summed bond distances approximately constant, with only a $0.5\%$ change across the low-pressure phase. 
The overall geometry of the interdimer H-bond network is also largely unchanged following the transition into the high-pressure $P 2_1 2_1 2_1$ phase, see Figures~\ref{Fig:CuV_inter_pdep}(c) and (d). 

Figure~\ref{Fig:CuV_intra_pdep}(a) shows the effect increasing pressure has on the intradimer bonds. Both the \chemfig{Cu-O=V} bond angle (top panel) and \chemfig{Cu-O=V} through-bond distance (bottom panel) decrease gradually by around $1.5\%$ as pressure is increased up to the structural transition. The intradimer Cu---O---H$\cdots$F---V bond distance (middle panel) is found to be relatively unchanged up to the point of the structural phase transition; again due to rotations of the water molecules.
At the transition to the high-pressure $P 2_1 2_1 2_1$ phase, the intradimer Cu---O---H$\cdots$F---V bond distance (middle panel) increases by $0.62$\,\AA. This is due to a rotation by $\approx 45^{\circ}$ of the VF$_4$ equatorial plane relative to the CuO$_4$ equatorial plane about the shared axial \chemfig{Cu-O=V} bond, as illustrated in Figures~\ref{Fig:CuV_intra_pdep}(b) and (c). This rotation causes a significant disruption of the intradimer hydrogen bonds, but is accompanied by only a small decrease in the \chemfig{Cu-O=V} bond angle ($\approx 5^{\circ}$), and the \chemfig{Cu-O=V} through-bond distance ($\approx 0.03(2)$\,\AA). 

\begin{figure*}[htb]
	\includegraphics[width=1.95\columnwidth]{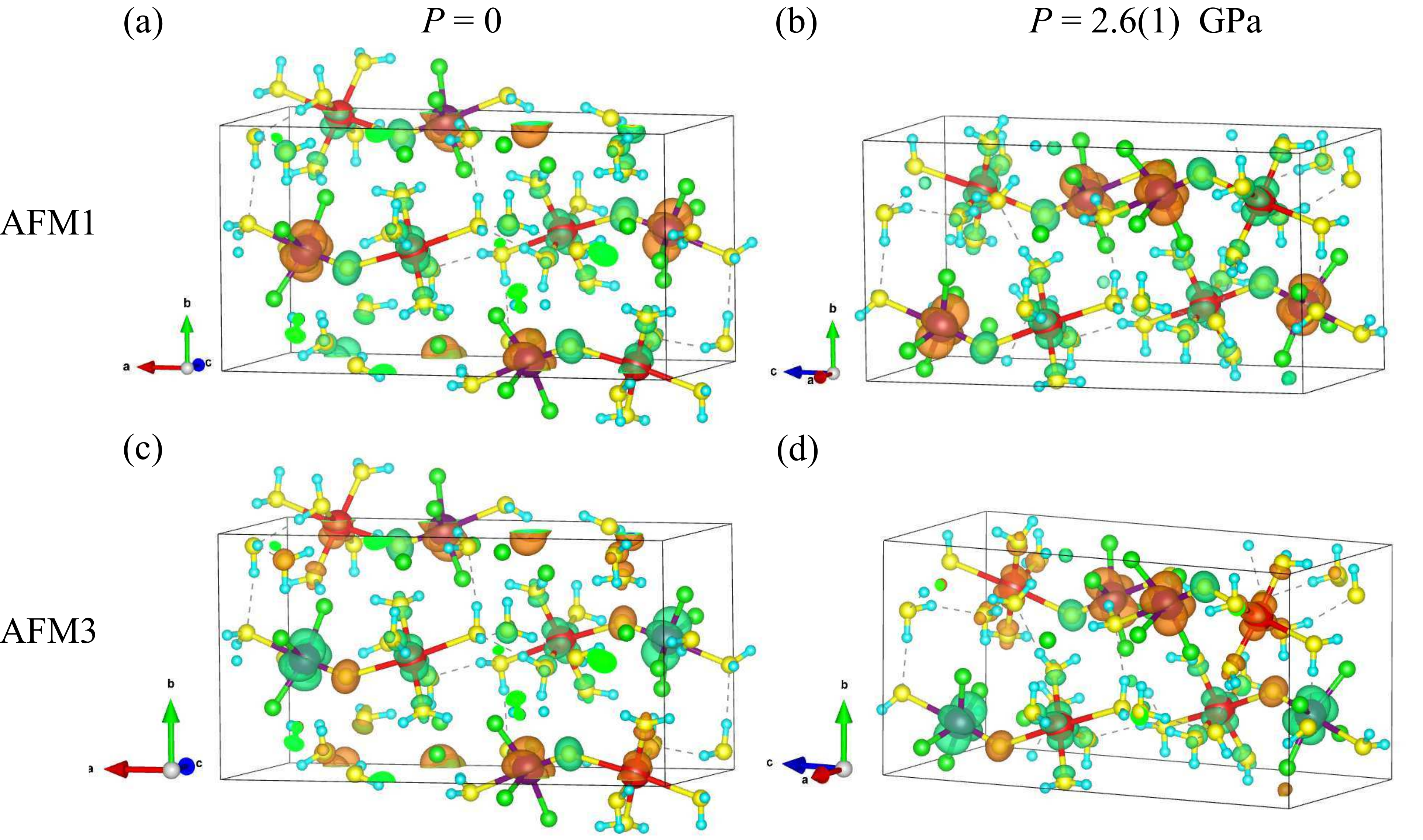}
	\caption{Spin-density distribution obtained from DFT for the two magnetic states that determine the strength of the intradimer exchange $J_0$. Orange and green isosurfaces represent regions with significant positive and negative spin density, respectively. These are shown for the ambient pressure structure in (a) and (c), and for the $p = 2.6(1)$~GPa structure, in (b) and (d).}
	\label{fig:spin_density}
\end{figure*}

\subsection{Density functional theory}

To help understand the differences in the strength of the magnetic exchange following the application of pressure, we used DFT to calculate the spin density distribution for the $p = 2.6$\,GPa structure. The DFT parameters used for these calculations are the same as those used for the ambient pressure structure, which can be found in the Supplemental Material of Ref.~\cite{Curley2021a}. Fig.~\ref{fig:spin_density} shows the spin-density distribution calculated for two different magnetic structures for both the ambient and 2.6\,GPa crystal structures, i.e. above and below the experimentally observed structural phase transition. In the magnetic structure labeled AFM1, all of the nearest-neighbor Cu--V couplings are antiferromagnetic, whereas in AFM3 the intradimer coupling is ferromagnetic, while all couplings between dimers are kept antiferromagnetic. These two magnetic states can be used in combination to calculate the magnitude of the intradimer exchange.

The antiferromagnetic state AFM1 is found to remain the magnetic ground state of the system at 2.6\,GPa. The spin-density distributions are very similar to those for the ambient pressure structure, with the most significant difference being changes in the direction in which the lobes of the spin density on the Cu and V ions point. The spin-density isosurfaces in the vicinity of Cu and V reflect the shapes of the magnetically active orbitals on these ions, whose directions move with the rotation of the equatorial ligands that accompanies the structural phase transition. Mulliken analysis reveals some changes in the total spin on the atoms after applying pressure. The moments on the Cu and V ions decrease to 0.62 and 1.21$\mu_\mathrm{B}$, respectively, from their values 0.65 and 1.29$\mu_\mathrm{B}$ for the ambient-pressure structure. Details of the moment sizes for each magnetic structure considered are given in the Appendix. The spin density on the O atoms joining Cu and V within a dimer, which we previously suggested to play a significant role in mediating the intradimer exchange, also decreases, from 0.26 to 0.21$\mu_\mathrm{B}$. Despite this, the relative energies of the magnetic states AFM1 and AFM3 leads to an estimate of $J_0 = 29.6$\,K for the intradimer exchange, a slight increase over the value $J_0 = 24.7(6)$\,K calculated for the ambient-pressure structure. While this is in qualitative agreement with experiment (see below), we also find that the calculated interdimer exchange $J'_{1–4}$ increases significantly, from 8.6 K to 33.0\,K, which is not reproduced in experiment. We note that, even at ambient pressure, DFT was found to significantly overestimate $J'_{1–4}$ which was found to be $\approx$ 1 K in experiment, with the DFT value also being very strongly dependent on the Hubbard $U$ used in the calculations \cite{Curley2021a}. This, coupled with the huge increase in $J'_{1–4}$ calculated for the high-pressure phase, suggests that the interdimer exchange $J'_{1–4}$ is poorly described by DFT. The other interdimer exchange $J''$ remains sufficiently small that it cannot be determined accurately.

We note that these results were obtained using a Hubbard $U$ of $5~\mathrm{eV}$ on the Cu and V $d$ orbitals, the value which was found to give the best agreement with experiment for the intradimer exchange $J_0$ in the ambient-pressure structure~\cite{Curley2021a}. However, it is possible that a different value of $U$ could result in better agreement with experiment for the $2.6~\mathrm{GPa}$ structure. Investigating the effect of Hubbard $U$ on the calculated exchange constants for the $2.6~\mathrm{GPa}$ structure, along with systematic investigations of the effect of pressure on the exchange constants obtained using DFT, are likely to be instructive but are beyond the scope of this study.

\subsection{High-pressure susceptometry}

\begin{figure}
\centering
\includegraphics[width= \linewidth]{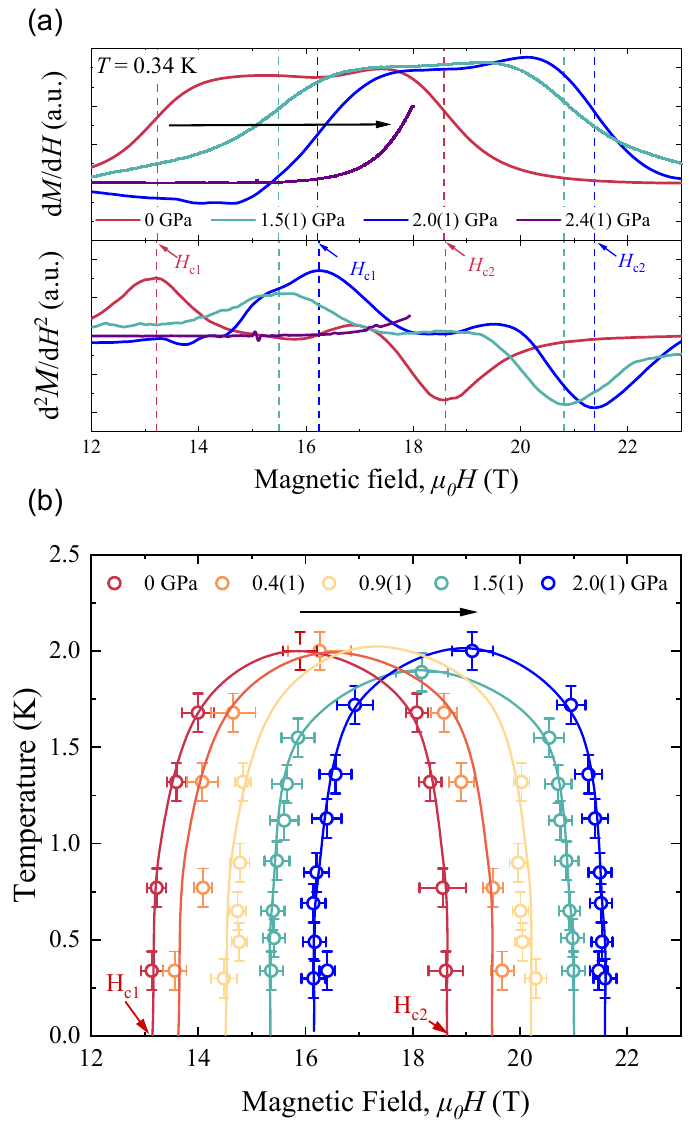}
\caption[width= \linewidth]{\small (a) Field-dependent differential susceptibility ${\rm d}M/{\rm d}H$ measured using radio-frequency susceptometry at the lowest temperature measured, at ambient pressure, 1.5(1) GPa, 2.0(1)\,GPa and 2.40(10)\,GPa. Critical fields at each pressure are determined from the positions of peaks and troughs in differentiated data (${\rm d}^2M/{\rm d}H^2$) and indicated with dashed lines. (b) Positions of the critical fields map out the pressure dependence of the temperature-field phase diagram of CuVOF$_4$(H$_2$O)$_6\cdot$H$_2$O.}  \label{Fig:CuV_BT_Pdep}
\end{figure}

The top panel of Figure~\ref{Fig:CuV_BT_Pdep}(a) shows the field dependence up to 23\,T of the differential susceptibility ${\rm d}M/{\rm d}H$ for a single crystal of CuVOF$_4$(H$_2$O)$_6\cdot$H$_2$O at the lowest temperature measured for each pressure, at ambient pressure, 1.5(1) GPa and 2.0(1)\,GPa, using the RF susceptometry technique
~\cite{Athas1993,Coffey2000,Clover1970}. The magnetic field was oriented close to parallel with the $a$ axis of the crystal. Typically in $S = 1/2$ dimer systems, sharp cusps are observed in ${\rm d}M/{\rm d}H$ at the critical fields $H_{\rm c1}$ and $H_{\rm c2}$~\cite{Lancaster2014b,Brambleby2017b} that respectively mark the transition from quantum disorder to field-induced magnetic order and then to the field-polarized phase. In the present case, the data instead exhibit two broad peak-like features similar to those observed in Ref.~\cite{Curley2021a}. It was suggested there that the broad nature of the features in ${\rm d}M/{\rm d}H$ observed at ambient pressure in this material could arise due to H-bond disorder within the complex interdimer exchange network, giving rise to a distribution in the superexchange between neighboring dimers and a smearing of the transition features in ${\rm d}M/{\rm d}H$.

The profile of the ${\rm d}M/{\rm d}H$ curves is qualitatively similar at low and high pressures, but the peak-like features are both shifted to higher fields at 2.0(1)\,GPa. The bottom panel shows the trough and peak features seen in the second differential of the magnetization, which can be used to determine the positions of $H_{\rm c1}$ and $H_{\rm c2}$, respectively~\cite{Curley2021a}. At 2.0(1)\,GPa, $H_{\rm c1} = 16.4(2)$\,T and $H_{\rm c2} = 21.5(2)$\,T, an increase of $\approx 3$\,T in both values compared to the ambient data. The difference between the two critical fields ($\Delta H = H_{\rm c2}-H_{\rm c1}$) at 2.0(1)\,GPa and 0.34\,T is $\Delta H = 5.1(3)$\,T, which is the same to within error as the ambient pressure value of $\Delta H = 5.5(3)$\,T; this is true for all five pressures at which susceptometry measurements were performed between ambient and 2.0(1)\,GPa.  At each pressure, ${\rm d}M/{\rm d}H$ was measured as a function of field at different temperatures for all pressures. The two peak-like features are seen to move closer with increasing temperatures, becoming unresolvable as separate features around 2\,K for all measured pressures. 

Also shown in Figure \ref{Fig:CuV_BT_Pdep}(a) are the susceptometry data taken at 2.4(1) GPa, which is higher than the critical pressure for the structural transition. Experimental constraints limit the field for this measurement to less than 18~T. This means that while the onset to the lower-field transition at $H_\mathrm{c1}$ is observed, the transition at $H_\mathrm{c2}$ is beyond the accessible field range. Nevertheless, it is seen that the slow increase in $H_\mathrm{c1}$ with pressure observed at lower pressures seems to be continued at 2.4(1) GPa. Certainly, the structural transition does not lead to a sudden drop in $H_\mathrm{c1}$ as might be expected if the intradimer exchange was mediated by the hydrogen-bond network, which is significantly disrupted at the structural transition.

The temperature dependence of the extracted $H_{\rm c1}$ and $H_{\rm c2}$ values allows the pressure-evolution of the temperature-field phase-diagram to be tracked, with the results shown in Figure~\ref{Fig:CuV_BT_Pdep}(b). A dome of field-induced $XY$ magnetic order is observed at all pressures, lying between the low-field quantum disordered and high-field field-polarized phases. The midpoint of the dome increases steadily with increasing pressure, but its width remains largely unchanged up to 2.0(1)\,GPa. This implies that the intradimer exchange strengthens under pressure, while the interdimer coupling stays roughly constant. Measurements of the pressure evolution of the $g$-factors are required to extract quantitative values for exchange energies. This process is discussed below.

\subsection{High-pressure electron spin resonance}

\begin{figure}
\centering\includegraphics[width=0.48\textwidth]{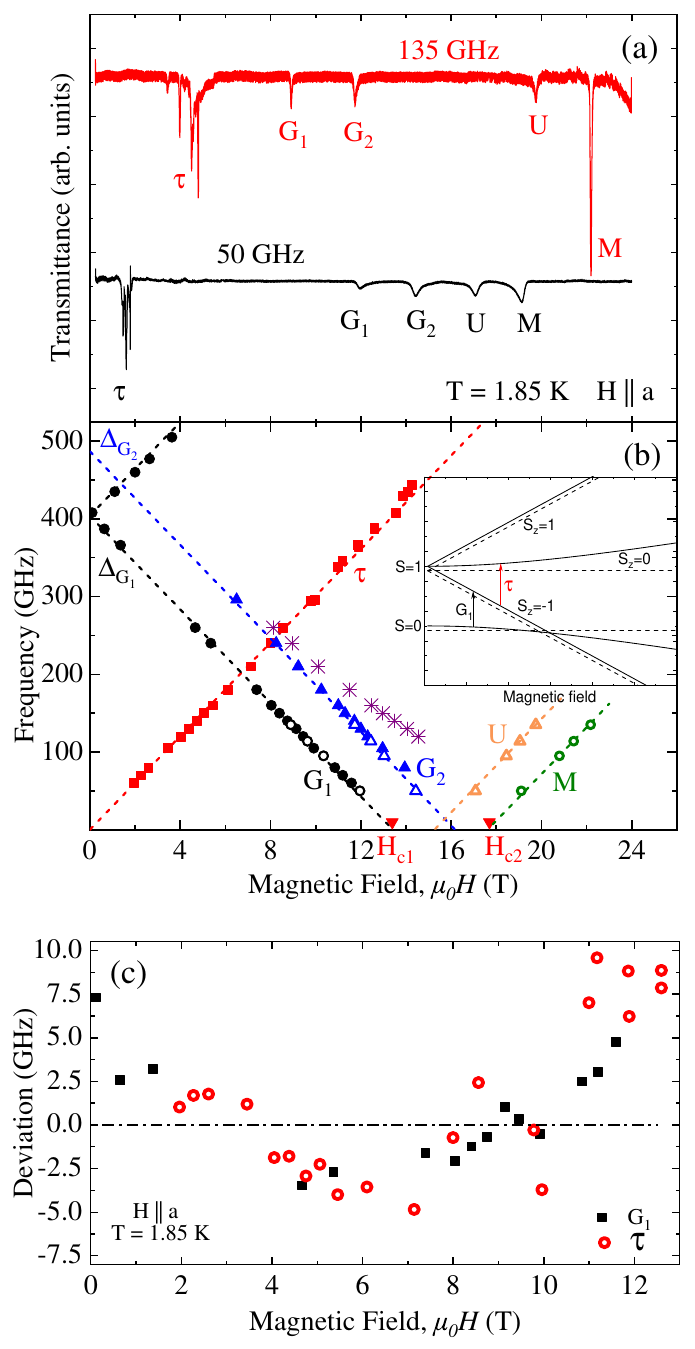}
	\caption{ Ambient pressure ESR data. (a) ESR spectra of CuVOF$_4$(H$_2$O)$_6$ measured at different radiation frequencies at 1.85\,K and $H \parallel a$. (b) Frequency-field dependence of the ESR transitions measured at 1.85\,K and $H \parallel a$. Red triangles show critical fields determined from ${\rm d}^2M/{\rm d}H^2$ data (Figure \ref{Fig:CuV_BT_Pdep}(b)). Closed symbols are the data from Ref~\cite{Curley2021a}, while open symbols show the extended dataset obtained at the 25\,T facility at Tohoku University. All observed resonances, including those marked with purple stars, are described in the text. The inset compares the energy levels of a dimer system composed of two ions with the same $g$-factor (dashed lines) with that of ions with different $g$-values (solid lines). Arrows indicate the modes $G_1$ and $\tau$ observed in the experiments. (c) Difference between the experimental data for modes $G_1$ (black) and $\tau$ (red) and their linear fits, with $\Delta{}G_1$=404 GHz and $g_{\mathrm{eff}}$ = 2.15.}
	
	\label{ESR_Fig1}   
\end{figure}

Here, we expand the previous ESR study of CuVOF$_4$(H$_2$O)$_6\cdot$H$_2$O \cite{Curley2021a} to encompass magnetic fields above $H_{\rm c2}$ and perform measurements at hydrostatic pressures up to 2.45(5) GPa. Figure \ref{ESR_Fig1}(a) shows the ambient pressure ESR spectra taken in a 25\,T superconducting magnet at 50 and 135\,GHz when the magnetic field is applied along the dimer axis ($H \parallel a$). We can identify five absorption peaks: $\tau$, G$_1$, G$_2$, M and U which originate from the Cu-V dimer system.

As shown previously by temperature dependence ESR measurements \cite{Curley2021a}, the $\tau$ mode corresponds to transitions ($\Delta S_z = \pm 1$) that occur within the excited $S = 1$ triplet state (see energy level diagram in inset of Fig.~\ref{ESR_Fig1}(b)). There are a few small peaks in the vicinity of $\tau$, which we attribute to low levels of Cu$^{2+}$ and V$^{4+}$ impurities \cite{Curley2021a}.

The frequency variation of the peak positions, the so-called frequency-field diagram (FFD), is shown in Fig.~\ref{ESR_Fig1}(b). The energy of the G$_1$ mode decreases with magnetic field and reaches zero close to $H_{\rm{c1}}$ determined from the RF susceptometry, indicating that it arises from transitions between the singlet ground state $S=0$ and the $S_z = - 1$ state. ESR measurements are sensitive to transitions at the $\Gamma$ point ($k=0$) because the radiation used ($\approx 500$~GHz) carries a negligible momentum $k$. Hence, the critical field at which the singlet-triplet energy gap closes extracted from ESR can be higher than that from bulk magnetometry, which is sensitive to the whole Brillouin zone. 

The slope of the dashed lines in Fig.~\ref{ESR_Fig1}(b) yields the effective $g$-factor of 2.15(1), which is the average value of the $g$-factors of the Cu$^{2+}$ and V$^{4+}$ ions that form the dimers. There is a clear departure from linear-in-field behavior of the $G_1$ and $\tau$ modes [see Fig.~\ref{ESR_Fig1}(c)]. This is a characteristic feature of a spin-dimer system formed by dissimilar spins. While the overall slope corresponds to the average $g$-factor of Cu$^{2+}$ and V$^{4+}$, the nonlinearity is dictated by the difference between them~\cite{Abragam1970}. 

The inset in Fig. \ref{ESR_Fig1}(b) illustrates the difference between the energy level schemes for dimer systems with similar (dashed lines) and dissimilar (solid lines) magnetic ions. The field behavior of the $S_z=0$ and $S = 0$ states is distinctly non-linear for the system of dissimilar spins; thus, the experimentally observed deviation from linearity allows us to disentangle the Cu$^{2+}$ and V$^{4+}$ $g$-factors. A~least squares fit returns $g_{\rm V}=2.00(5)$ and $g_{\rm Cu}=2.30(5)$. We note that the presence of DM interaction may also lead to the non-linear behavior of the $G_1$ mode, depending on the relative orientation of the DM vector. However, the deviation from the linearity in the case of DM interaction is much more pronounced in the vicinity of the $H_{c1}$, while here we observe it over the entire field range~\cite{Nojiri2003}.  

The $M$ mode appears close to $H_{\rm c2}$ and grows linearly with increasing field. It is attributed to the excitation of magnons emerging from the field-polarized phase. Following the same Brillouin-zone argument above, the field at which this mode reaches zero frequency can be lower than the upper critical field obtained from bulk measurements. 

\begin{figure}
	\includegraphics[width=0.42\textwidth]{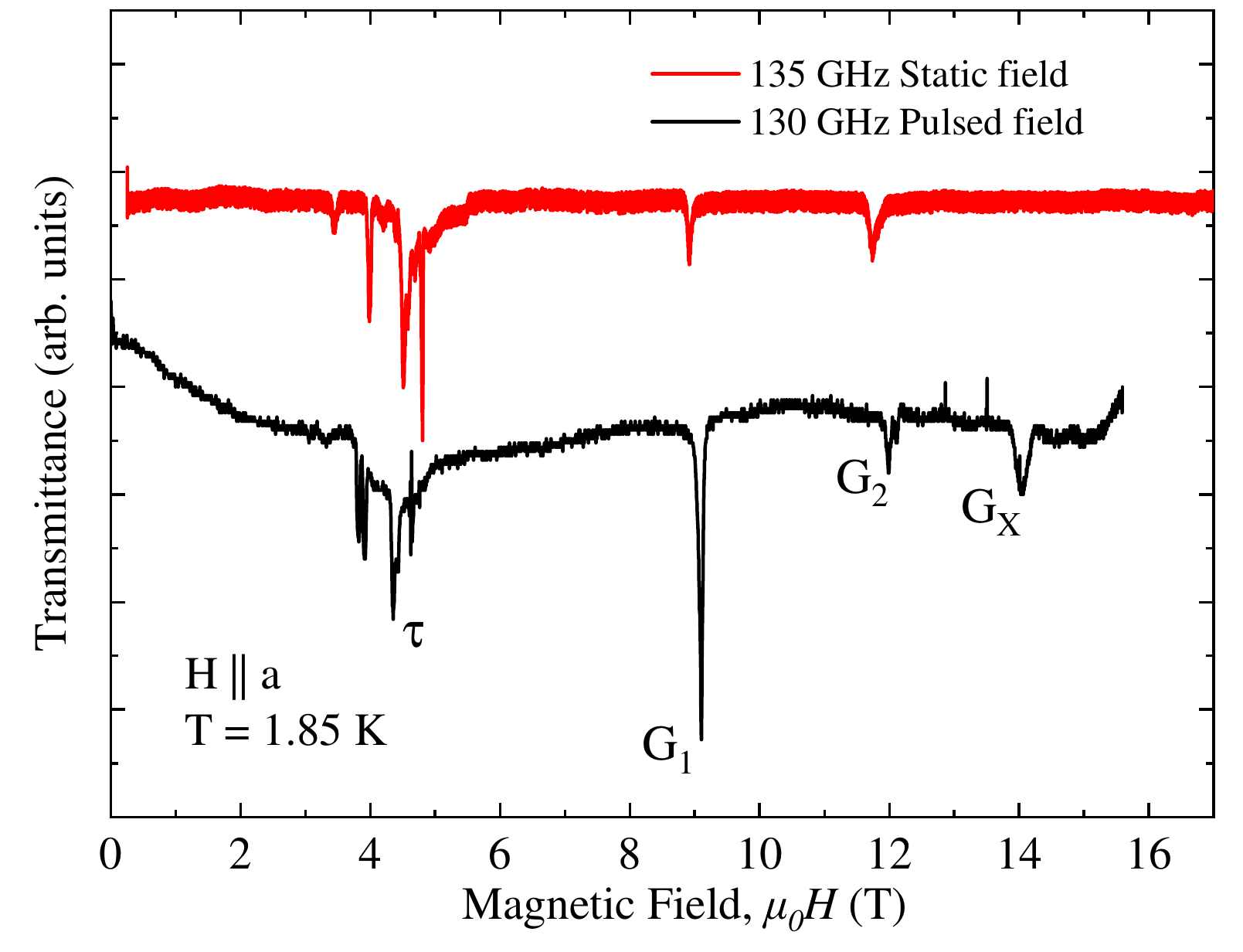}
	\caption{Ambient pressure, 1.85\,K ESR spectra of CuVOF$_4$(H$_2$O)$_6$ measured in pulsed (black line) and quasi-static magnetic fields (red line) at 130 and 135 GHz, respectively, for $H\parallel a$.}
		\label{Pulsed_Static}   
\end{figure}

\begin{figure}
	\includegraphics[width=0.38\textwidth]{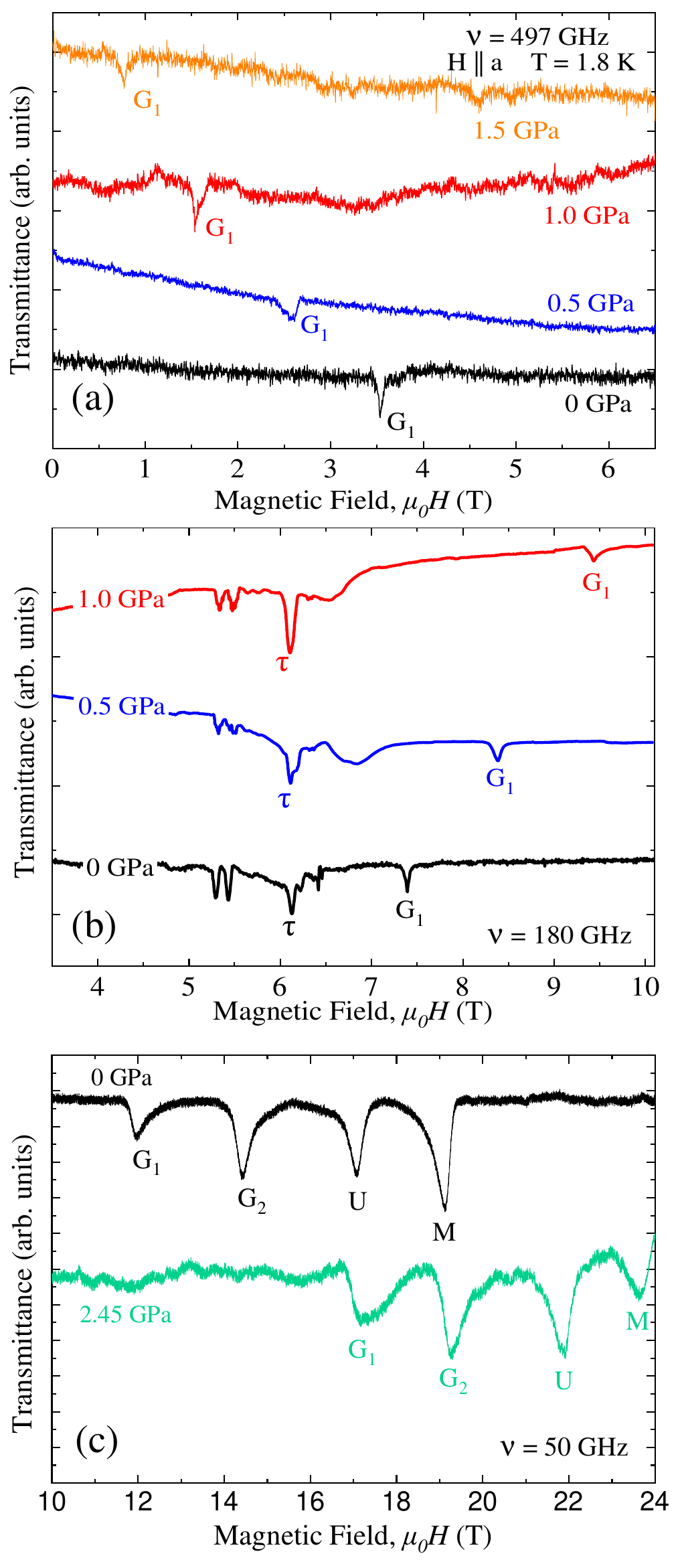}
	\caption{Example 497\,GHz (a), 180\,GHz (b), and 50\,GHz (c) ESR spectra of CuVOF$_4$(H$_2$O)$_6$ measured at different hydrostatic pressures, $T = 1.8$\,K and $H \parallel a$. The spectra at 50\,GHz and 2.45(5)\,GPa (shown in green on panel c) are measured above the critical pressure for the structural transformation.}
		\label{pres_a}   
\end{figure}

The triplet excitations in a system of equivalent dimers appear in the ESR spectrum as a single gapped mode. Additional $G_n$ modes in the spectra with the same slope as $G_1$ have therefore previously been ascribed to the presence of more than one nonequivalent dimer systems in the unit cell, leading to the formation of additional triplet branches 
\cite{Leuenberger1984, Matsumoto2008, Kofu2009a, Kofu2009, Wang2014, Kamenskyi2013}. 
Following this example, we attribute the $G_2$ mode to the excitations of a second triplet branch. 
Transitions indicated by purple stars in Fig.~\ref{ESR_Fig1}(b) are only observed in the pulsed-field ESR experiments and are absent in quasi-static field measurements (see Fig. \ref{Pulsed_Static}). This mode has a pronounced non-linear magnetic field behavior and appears in the vicinity of $H_{\rm{c1}}$. Although the exact origin of these excitations is unclear at the moment, we suspect that it may appear due to magnetostrictive and magnetocaloric effects, which typically become quite strong close to the critical field. These effects have been observed before in studies of dimer systems, particularly in the high ${\rm d}H/{\rm d}t$ provided by pulsed magnets~\cite{Nomura2020,Brambleby2017b}. 

The intensity of the U mode observed at fields between $G_2$ and M decays as the frequency grows and the resonance shifts beyond $H_{\rm c2}$ (shown with orange symbols in Fig.~\ref{ESR_Fig1}(b)). The excitation spectrum in the XY-ordered phase is formed by the excited triplons and the intricate network of interactions between them and is known to support multiple ESR resonances. For example, very complicated ESR spectra between $H_{\rm c1}$ and $H_{\rm c2}$ have been reported in Ba$_3$Cr$_3$O$_8$~\cite{Kamenskyi2013} and SrCu$_2$(BO$_3$)$_2$~\cite{Nojiri2003,Nojiri1999}.

Figure \ref{pres_a} illustrates the changes in the ESR spectra caused by hydrostatic pressure for $H \parallel a$, measured at frequencies both above and below the zero-field singlet-triplet gap, $\Delta_{G_1} \approx 400$\, GHz at ambient pressure. For frequencies above the gap ({panel (a), $\nu=497$\,GHz) , the $G_1$ mode shifts towards lower magnetic fields with increasing pressure, while below the gap (panel (b), $\nu = 180$\,GHz) it shifts to higher fields, which suggests that $\Delta_{G_1}$ increases with pressure. Panel (c) compares high-field ESR spectra at ambient and above critical pressure. The spectra remain qualitatively unchanged by the structural transformation but the resonances are shifted to higher field values at the higher pressure.

The pressure also affects the $g$-factor through modifications to the local spin environment. The constraints of the pressure experiments means that is not possible to disentangle the Cu$^{2+}$ and V$^{4+}$ $g$-factors individually at elevated pressures with acceptable accuracy, but the effective $g$ factor for the dimer unit can be ascertained. The parameters extracted from the ESR data are discussed in detail below in comparison to the susceptometry results.

\section{Discussion}

\begin{figure}
\centering
\includegraphics[width= \linewidth]{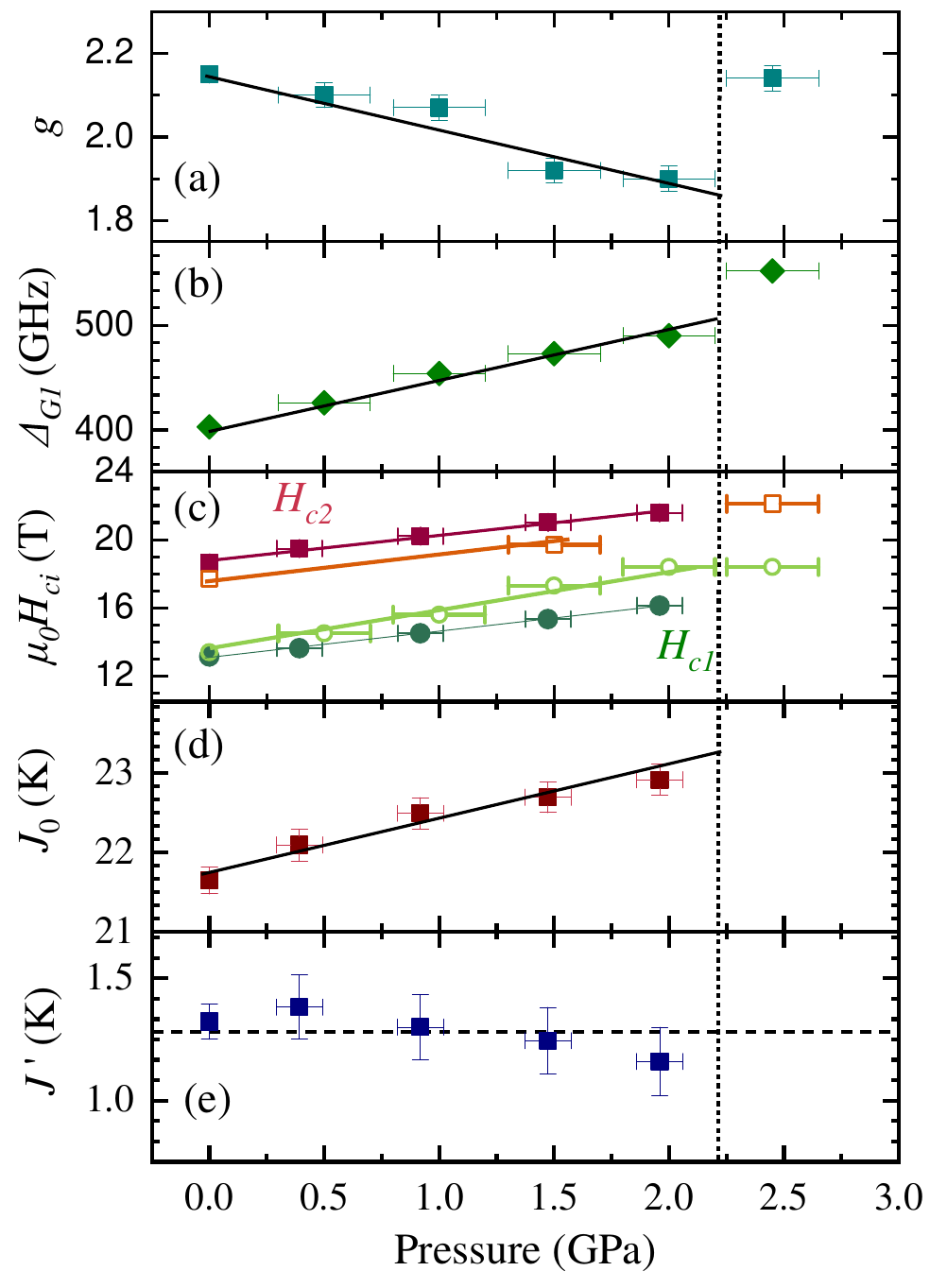}
\caption[width= \linewidth]{\small Pressure dependence of the magnetic parameters. (a) Effective $g$ factor extracted from ESR. (b) Triplet-singlet energy gap, $\Delta_{G1}$, extracted from ESR. (c) 
Values of critical fields $H_{\rm c1}$ and $H_{\rm c2}$, found from extrapolating the susceptometry-derived data in Fig. \ref{Fig:CuV_BT_Pdep} to zero temperature (solid symbols) and from the zero-frequency intercepts of the $G_1$ and $M$ ESR modes (open symbols). (d) Values of intradimer exchange $J_0$ and (e) effective interdimer exchange $J'$ calculated from  Eqn~\ref{eq:Tachiki}. The vertical dashed line indicates the pressure-induced structural phase transition.
}   
\label{Fig:Discussion_ParamsFigure}
\end{figure}

Fig.~\ref{Fig:Discussion_ParamsFigure} displays the effect of hydrostatic pressure on the magnetic parameters of CuVOF$_4$(H$_2$O)$_6$$\cdot$H$_2$O for magnetic fields applied along $a$. The vertical dashed line in the figure indicates the critical pressure at which takes place the structural phase transition described above. Panel (a) shows the effective dimer $g$ factor extracted from ESR. A gradual monotonic decrease is observed on increasing pressure from 2.15(1) at ambient pressure to 1.90(3) at 2.0(2)\,GPa. At the structural transition, the local spin environment undergoes a significant reconfiguration and the $g$ factor measured at 2.45(5)\,GPa increases sharply to 2.14(3), the same as the ambient pressure value within experimental uncertainty. The zero-field singlet-triplet gap, $\Delta_{G1}$, increases linearly with pressure, see panel (b). However, unlike the $g$ factor, this parameter shows only a small deviation from this linear behavior on passing through the structural transition. 

Fig.~\ref{Fig:Discussion_ParamsFigure}(c) shows the pressure dependence of the critical fields $\mu_0 H_{\rm c1}$ and $\mu_0 H_{\rm c2}$ extracted from RF susceptometry measurements by extrapolating the finite-temperature critical-field values to zero temperature (solid symbols). Respectively, these are the end of the quantum-disordered phase and the onset of the field-saturated phases and hence demark the zero-temperature bounds of the dome of field-induced $XY$-order shown in Fig.~\ref{Fig:CuV_BT_Pdep}. As can be seen, both critical fields increase linearly with pressure at roughly the same rate, indicating that, as mentioned earlier, the dome moves higher in field with applied pressure, but does not change in width. Also shown in panel (c) are the critical fields as deduced from ESR measurements (open symbols). In this case, the lower and upper critical fields are defined as the field at which the $G_1$ and $M$ ESR modes reach zero frequency, respectively. The $M$ mode was only observed in a subset of the pressure measurements, those where high fields were employed. Given the fact, discussed above, that ESR is sensitive only to $k = 0$ transitions, these fields need not be identical to, but should lie within those extracted from susceptometry. This is what we observed and, in fact, there is a reasonably close agreement between the critical fields determined from the two techniques. 

Within the model of weakly interacting dimers, the critical fields are determined by the exchange energies via the relations~\cite{Tachiki1970} 
\begin{equation}
    \label{eq:Tachiki}
    g \mu_{\rm{B}} \mu_0 H_{\rm{c1}} = J_0-nJ'/2, {\rm ~~~~}
    g \mu_{\rm{B}} \mu_0 H_{\rm{c2}} = J_0+nJ'. 
\end{equation}
Here $J_0$ is the primary, intradimer exchange defined in Eq.~\ref{eq:Hamiltonian}, $J'$ is the effective interdimer interaction strength (the average of $J'_{1–4}$) and $n = 4$ is the number of interdimer neighbours as described above. This is the same model previously used to describe the ambient pressure properties of CuVOF$_4$(H$_2$O)$_6$$\cdot$H$_2$O~\cite{Curley2021a}. Using the critical fields extracted from RF susceptometry and the $g$ factors from ESR, the pressure dependence of the exchange energies can be determined and are shown in Fig.~\ref{Fig:Discussion_ParamsFigure}(d) and (e). At ambient pressure, both interaction strengths are found to be the same as reported in Ref.~\cite{Curley2021a}. 

Up to 2.0(1)\,GPa, pressure has little effect on the magnitude of the interdimer exchange $J'$, which remains constant within experimental uncertainty at $J' \approx 1.3(1)$\,K. This is in line with high-pressure structural measurements, which show that whilst pressure does bring the dimer-units into closer proximity, the overall geometry and length of the interdimer exchange pathways is nearly unchanged up to 3.0~GPa; see Figure~\ref{Fig:CuV_inter_pdep}. In contrast, the magnitude of $J_0$ is seen to increase linearly with pressure, reaching a maximum measured value of 22.9(2)\,K at 2.0(1)\,GPa within the low-pressure structural phase. This increase is due to decreases in both the Cu---O---V bond length and angle (see Figure~\ref{Fig:CuV_intra_pdep}(a)) and is in agreement with the trend found by the DFT calculations.   

It is not possible to follow the trends in $J_0$ and $J'$ through the pressure-induced structural phase transition because $H_{\rm c2}$ grows larger than the available magnetic field. However, the singlet-triplet gap determined from ESR [panel (b)] shows only a small deviation at the structural transition from the slow increase seen within the low-pressure phase, indicating that there is no radical change in the size of $J_0$ at the structural phase transition. This observation provides additional strong evidence for the role of intradimer oxygen in the exchange pathway. As discussed above, the summed intradimer Cu---O---V bond distance decreases only slightly across the transition, and the bond-angle reduces by $\approx 5^{\circ}$ when moving from 0 to 2.0(1)\,GPa [Figure~\ref{Fig:CuV_intra_pdep}(a)]. 
On the other hand, the structural phase-transition induces a rotation of the equatorial ligands on each ion about the Cu---O---V bond, see Figures~\ref{Fig:CuV_intra_pdep}(b) and \ref{Fig:CuV_intra_pdep}(c). This is reflected in the significant change in the $g$ factor at the transition [Fig.~\ref{Fig:Discussion_ParamsFigure}(a)]. The rotation causes a drastic increase in the length of the summed intradimer Cu---O---H$\,\cdots$F---V bond pathway by $0.62(2)$\,\AA\ at the onset of the phase transition. If this were the primary exchange pathway, then a significant change would be expected in the intradimer exchange and hence also the singlet-triplet gap, which is not observed. Instead, the strong $J_0$ must be accounted for via the highly unusual mechanism suggested in Ref~\cite{Curley2021a} in which the intradimer superexchange is mediated along the Cu(II) JT axis to V(IV) via a single vanadyl oxygen bridge.

In summary, we find that the primary exchange energy in the quantum magnet CuVOF$_4$(H$_2$O)$_6$$\cdot$H$_2$O is enhanced by applied hydrostatic pressure up to and across a structural phase transition at around 2.4\,GPa, while the interdimer exchange changes only slightly. This is consistent with the pressure-induced modifications observed in the bond distances and the $g$ factors. The effect of the dissimilarity of the spins in the dimer unit is seen in the ESR transition energies, which exhibit a non-linear field dependence, as well as several unidentified resonances. It remains to be seen if other spin-1/2 copper-vanadium compounds can be identified that host exchange interactions mediated through the oxygen atom on the distorted JT axes.

\section{Acknowledgments}
We thank J. Villa for useful discussions, as well as T. Orton and P. Ruddy for technical assistance. This project has received funding from the European Research Council (ERC) under the European Union’s Horizon 2020 research and innovation programme (Grant Agreement No. 681260). A portion of this work was performed at the National High Magnetic Field Laboratory (NHMFL), which is supported by National Science Foundation Cooperative Agreement Nos. DMR-1644779 and DMR-2128556 and the Department of Energy (DOE). MJC is supported by a UKRI Future Leaders Fellowship, Grant No. MR/Y016602/1. TL is supported by EPSRC (UK) through grant no. EP/Z534067/1. This work has made use of the Hamilton HPC Service of Durham University. The work at EWU was supported by the NSF through grant no. DMR-2104167. High-field ESR experiments were performed at the High Field Laboratory for Superconducting Materials, Institute for Materials Research, Tohoku University (proposals 20K0603  and 20H0507). Support of the ICC-IMR Visitor Program at Tohoku University is acknowledged. T. S. acknowledges the support provided by the JSPS KAKENHI Grant Numbers 23K03321. D. K. was supported by the PRIME program of the German Academic Exchange Service (DAAD) with funds from the German Federal Ministry of Education and Research (BMBF). D.K and A.C. thank the Deutsche Forschungsgemeinschaft (DFG, German Research Foundation)–TRR 360–492547816 for supporting this work. 

Data presented in this paper are available for download at \href{https://wrap.warwick.ac.uk/200652/}{https://wrap.warwick.ac.uk/200652/} \cite{DataRepo_Placeholder}. 

For the purpose of open access, the author has applied a Creative Commons Attribution (CC-BY) license to any Author Accepted Manuscript version arising from this submission.

\appendix*
\section{DFT Calculations - additional details on magnetic moments of each spin configuration}

\begin{table}[!]
\caption{DFT calculations details - Local magnetic moments (in Bohr magnetons) on Cu and V for the FM and AFM configurations. Moments are listed separately for ions with positive spin density and those with negative spin density.}
\label{tab:moments}
\begin{ruledtabular}
\begin{tabular}{lcccc}
Configuration & \multicolumn{2}{c}{Cu} & \multicolumn{2}{c}{V} \\
\cline{2-3}\cline{4-5}
& + & - & + & - \\
\hline
FM   & 0.622 & 0.622 & 1.198 & 1.198 \\
AFM1 & 0.624 & 0.624 & 1.205 & 1.205 \\
AFM2 & 0.626 & 0.625 & 1.208 & 1.207 \\
AFM3 & 0.610 & 0.617 & 1.196 & 1.197 \\
AFM4 & 0.618 & 0.617 & 1.199 & 1.199 \\
AFM5 & 0.619 & 0.618 & 1.206 & 1.206 \\
\end{tabular}
\end{ruledtabular}
\end{table}

For the DFT results presented in the main text, the moments on each of the ions in each spin configuration were calculated using Mulliken population analysis, and the results are shown in Table~\ref{tab:moments}. We see that the variation in the moment size across configurations is small, with the total variation in the Cu moment (range divided by mean) being around 3\%, while the V moments are even more consistent, with these agreeing to within 1\%. For some antiferromagnetic (AFM) configurations, we calculated a slightly different moment for those ions having a net positive spin density compared to those having a negative spin density, which might suggest that some of these AFM configurations have net magnetization. However, in these cases, the total moment of the unit cell is $<2\times10^{-5}$ $\mu_\mathrm{B}$, indicating that the spin density on the other atoms in the system compensates for this imbalance.

\providecommand{\noopsort}[1]{}\providecommand{\singleletter}[1]{#1}%

\end{document}